\documentclass{PoS}

\usepackage{alltt,axodraw,graphicx}

\newcommand{\beq}{\begin{equation}}
\newcommand{\eeq}{\end{equation}}
\newcommand{\ice}[1]{\relax}

\title{Parallel versions of the symbolic manipulation system FORM}
\ShortTitle{Parallel FORM}

\author{\speaker{M.~Tentyukov}\\
        Institut f\"ur Theoretische Teilchenphysik, Karlsruhe
        Institute of Technology (KIT), D-76128 Karlsruhe, Germany\\
        E-mail: \email{tentukov@particle.uni-karlsruhe.de}}
\author{
        J.A.M.~Vermaseren \\ 
        Nikhef Science Park 105 1098 XG, Amsterdam\\
        E-mail:\email{t68@nikhef.nl}}
\author{
        J.~Vollinga \\ 
        Nikhef Science Park 105 1098 XG, Amsterdam\\
        E-mail:\email{jensv@nikhef.nl}}

\abstract{
The symbolic manipulation program FORM is specialized to handle very
large algebraic expressions.  Some specific features of its internal
structure make FORM very well suited for parallelization.

We have now two parallel versions of FORM, one is based on POSIX threads
and is optimal for modern multicore computers while another one uses
MPI and can be used to parallelize FORM on clusters and Massive
Parallel Processing systems. Most existing FORM programs will be able
to take advantage of the parallel execution without the need for
modifications.
}

\FullConference{13th International Workshop on Advanced Computing and
  Analysis Techniques in Physics Research\\
                 February 22-27, 2010\\
                 Jaipur, India}

\begin{document}
\section{Introduction}
The symbolic manipulation system FORM~\cite{FORM} which
is available already more than 20 years, is specialized to handle very large algebraic expressions of billions of terms in an efficient and
reliable way. It is widely used, in particular in the
framework of perturbative Quantum Field Theory, where sometimes hundreds of
thousands of Feynman diagrams have to be computed; most of the 
spectacular calculations of refs~\cite{FORMused,parFORMused}
would hardly have been possible with other available systems.  However, the
abilities of FORM are also quite useful in other fields of science
where the manipulation of huge expressions is necessary.

Parallelization is one of the most efficient ways to increase
performance. Some internal specifics \cite{ParFORM}
make FORM very well suitable for parallelization so the idea to
parallelize FORM is quite natural.

\section{General concepts and models in use}
The general concept of FORM parallelization is as follows \cite{ParFORM,ACAT08,TFORM}:
upon the startup, the program launches a {\em master} and several
{\em workers}. FORM treats each expression individually, which allows
the master to split incoming expressions into independent chunks. Each
chunk is processed by workers in parallel, and then the master
collects the results.

At present, we have two different models~\cite{ACAT08,TFORM}:
in {\tt ParFORM} \cite{ParFORM} the master and workers are independent processes
communicating via MPI\footnote{A Message Passing Interface, see http://www.mpi-forum.org/} 
and in {\tt TFORM}~\cite{TFORM} master and workers are separate
threads\footnote{TFORM uses POSIX threads, or pthreads} of
a multithreaded process. 

Both models require almost no special efforts for
parallel programming, all FORM programs may be executed in parallel
without any changings. The user may give FORM some hints of how to
parallelize some things better; these hints are simply ignored by the
sequential version of FORM.

Since TFORM uses common address space, it is runnable only on SMP
computers. On the other hand, sometimes it permits more efficient
parallelization, and it does not depend on MPI which make it much easier
for deployment. ParFORM can be used not only on SMP computers but also
in clusters and Massive Parallel Processors (MPP).

\begin{figure}[ht]
\begin{center}
\begin{picture}(370,150)(0,0)
\SetOffset(35,15)
\LogAxis(0,0)(320,0)(1.6,3,0,2)
\LogAxis(0,125)(320,125)(1.6,-3,0,2)
\LogAxis(0,0)(0,125)(1,-3,63,2)
\LogAxis(320,0)(320,125)(1,3,63,2)
\Text(-5,13)[r]{8000}
\Text(-5,25)[r]{10000}
\Text(-5,84)[r]{30000}
\Text(-5,100)[r]{40000}
\Text(-5,112)[r]{50000}
\Text(60,-5)[t]{2}
\Text(95,-5)[t]{3}
\Text(120,-5.0)[t]{4}
\Text(181,-5.0)[t]{8}
\Text(200,-5.0)[t]{10}
\Text(260,-5.0)[t]{20}
\Text(293,-5.0)[t]{30}
\rText(-20,56)[][l]{Time (s)}
\Text(150,-5)[t]{Workers}
\Line(60.205999,98.004082)(95.424251,74.962395)
\Line(95.424251,74.962395)(120.411998,56.598045)
\Line(120.411998,56.598045)(139.794001,47.926918)
\Line(139.794001,47.926918)(155.630250,36.482892)
\Line(155.630250,36.482892)(169.019608,26.517724)
\Line(169.019608,26.517724)(180.617997,20.330374)
\Line(180.617997,20.330374)(190.848502,18.989706)
\Line(190.848502,18.989706)(200.000000,17.262467)
\Line(200.000000,17.262467)(208.278537,15.659691)
\Line(208.278537,15.659691)(215.836249,14.537263)
\Line(215.836249,14.537263)(222.788670,13.700253)
\Line(222.788670,13.700253)(229.225607,13.231161)
\Line(229.225607,13.231161)(235.218252,12.057096)
\Line(235.218252,12.057096)(240.823997,11.271598)
\Line(240.823997,11.271598)(246.089784,10.504637)
\Line(246.089784,10.504637)(251.054501,12.630226)
\Line(251.054501,12.630226)(260.205999,10.772903)
\Line(260.205999,10.772903)(276.042248,10.721234)
\Line(276.042248,10.721234)(289.431606,11.395402)
\Line(289.431606,11.395402)(301.029996,11.641125)
\Vertex(60.205999,98.004082){2.5}
\Vertex(95.424251,74.962395){2.5}
\Vertex(120.411998,56.598045){2.5}
\Vertex(139.794001,47.926918){2.5}
\Vertex(155.630250,36.482892){2.5}
\Vertex(169.019608,26.517724){2.5}
\Vertex(180.617997,20.330374){2.5}
\Vertex(190.848502,18.989706){2.5}
\Vertex(200.000000,17.262467){2.5}
\Vertex(208.278537,15.659691){2.5}
\Vertex(215.836249,14.537263){2.5}
\Vertex(222.788670,13.700253){2.5}
\Vertex(229.225607,13.231161){2.5}
\Vertex(235.218252,12.057096){2.5}
\Vertex(240.823997,11.271598){2.5}
\Vertex(246.089784,10.504637){2.5}
\Vertex(251.054501,12.630226){2.5}
\Vertex(260.205999,10.772903){2.5}
\Vertex(276.042248,10.721234){2.5}
\Vertex(289.431606,11.395402){2.5}
\Vertex(301.029996,11.641125){2.5}
\end{picture}
\caption{\label{MZVDesy}
Running times of the Multiple Zeta Value TFORM program. 
The runs were for weight 23, up to depth 7.
}
\end{center}
\end{figure}
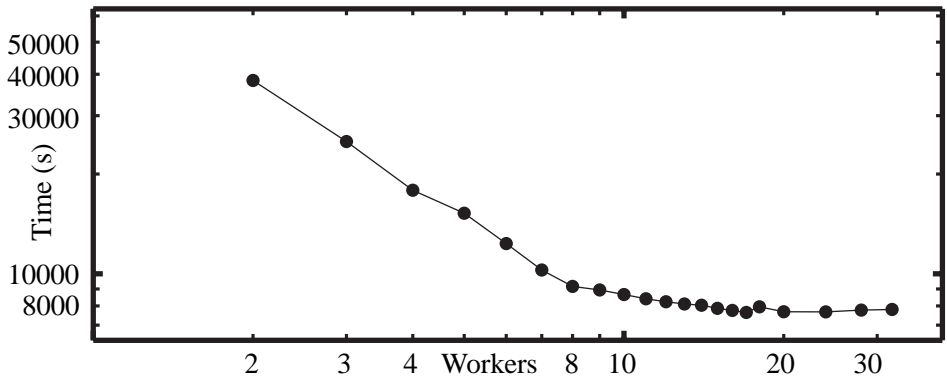

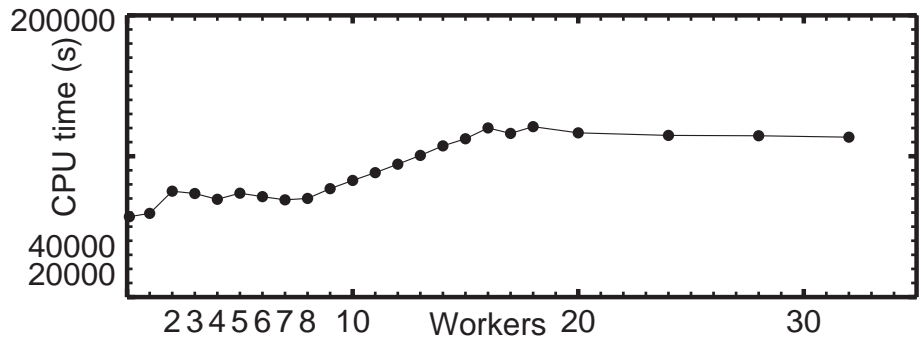
\begin{figure}[ht]
\begin{center}
\begin{picture}(360,120)(0,0)
\SetOffset(50.000000,15.000000)
\SetScale{0.85}
\SetPFont{Helvetica}{14}
\LinAxis(0.000000,0.000000)(350.000000,0.000000)(3.5,10,3,0,2)
\LinAxis(0.000000,125.000000)(350.000000,125.000000)(3.5,10,-3,0,2)
\LinAxis(0.000000,0.000000)(0.000000,125.000000)(2,10,-3,0,2)
\LinAxis(350.000000,0.000000)(350.000000,125.000000)(2,10,3,0,2)
\PText(-5.000000,12.500000)(0)[r]{20000}
\PText(-5.000000,25.000000)(0)[r]{40000}
\PText(-5.000000,125.000000)(0)[r]{200000}
\PText(20.000000,-1.500000)(0)[t]{2}
\PText(30.000000,-1.500000)(0)[t]{3}
\PText(40.000000,-1.500000)(0)[t]{4}
\PText(50.000000,-1.500000)(0)[t]{5}
\PText(60.000000,-1.500000)(0)[t]{6}
\PText(70.000000,-1.500000)(0)[t]{7}
\PText(80.000000,-1.500000)(0)[t]{8}
\PText(100.000000,-1.500000)(0)[t]{10}
\PText(200.000000,-1.500000)(0)[t]{20}
\PText(300.000000,-1.500000)(0)[t]{30}
\PText(-30.000000,75.750000)(90)[]{CPU time (s)}
\PText(160.000000,-10.000000)(0)[]{Workers}
\Line(1.000000,35.673819)(10.000000,37.138481)
\Line(10.000000,37.138481)(20.000000,46.989838)
\Line(20.000000,46.989838)(30.000000,45.952832)
\Line(30.000000,45.952832)(40.000000,43.390563)
\Line(40.000000,43.390563)(50.000000,46.095919)
\Line(50.000000,46.095919)(60.000000,44.563575)
\Line(60.000000,44.563575)(70.000000,43.147250)
\Line(70.000000,43.147250)(80.000000,43.774225)
\Line(80.000000,43.774225)(90.000000,48.108744)
\Line(90.000000,48.108744)(100.000000,51.754731)
\Line(100.000000,51.754731)(110.000000,55.203837)
\Line(110.000000,55.203837)(120.000000,58.960400)
\Line(120.000000,58.960400)(130.000000,62.864187)
\Line(130.000000,62.864187)(140.000000,67.089207)
\Line(140.000000,67.089207)(150.000000,70.248619)
\Line(150.000000,70.248619)(160.000000,75.048525)
\Line(160.000000,75.048525)(170.000000,72.621032)
\Line(170.000000,72.621032)(180.000000,75.616681)
\Line(180.000000,75.616681)(200.000000,72.867706)
\Line(200.000000,72.867706)(240.000000,71.738418)
\Line(240.000000,71.738418)(280.000000,71.577163)
\Line(280.000000,71.577163)(320.000000,70.954281)
\Vertex(1.000000,35.673819){2.5}
\Vertex(10.000000,37.138481){2.5}
\Vertex(20.000000,46.989838){2.5}
\Vertex(30.000000,45.952832){2.5}
\Vertex(40.000000,43.390563){2.5}
\Vertex(50.000000,46.095919){2.5}
\Vertex(60.000000,44.563575){2.5}
\Vertex(70.000000,43.147250){2.5}
\Vertex(80.000000,43.774225){2.5}
\Vertex(90.000000,48.108744){2.5}
\Vertex(100.000000,51.754731){2.5}
\Vertex(110.000000,55.203837){2.5}
\Vertex(120.000000,58.960400){2.5}
\Vertex(130.000000,62.864187){2.5}
\Vertex(140.000000,67.089207){2.5}
\Vertex(150.000000,70.248619){2.5}
\Vertex(160.000000,75.048525){2.5}
\Vertex(170.000000,72.621032){2.5}
\Vertex(180.000000,75.616681){2.5}
\Vertex(200.000000,72.867706){2.5}
\Vertex(240.000000,71.738418){2.5}
\Vertex(280.000000,71.577163){2.5}
\Vertex(320.000000,70.954281){2.5}
\end{picture}
\caption{\label{CPUtime}
Total CPU time of the Multiple Zeta Value TFORM program.
}
\end{center}
\end{figure}

\section{Performance}
Both ParFORM and TFORM demonstrate approximately the same
speedup~\cite{ACAT08,TFORM}. Here we discuss TFORM running the Multiple
Zeta Value program~\cite{MZV} on the computer ``qftquad5'' at DESY. The
computer has 96 GB of main memory and 8 independent CPU cores; the
effective number of CPU cores is 16 due to hyperthreading. The results
are given in Fig.~\ref{MZVDesy}.

For reference, the run with FORM (the sequential version) took 57078 sec.

We see three regions: first, the speedup is almost linear up to 8
workers; second, the speedup is also almost linear in the range of 8-16
workers but with much less slope, and after 16 workers we observe a
saturation. When we looked at the total  amount of CPU time used,
Fig.\ref{CPUtime}, we see the total CPU time is more or less constant 
up to 8 workers and  above 16 workers. In the range of 8-16 workers however it increases 
steadily. This is responsible for the slower decline in real time in the 
first graph, because the pseudo efficiency (total CPU time divided by real 
time and divided by number of workers) remains more or less the same in 
this range. This is behaviour that is typical for hyperthreading.
The total amount of work that can be 
obtained from this computer is about 9.5 times the amount that can be 
obtained from a single core.

The analysis of the data reveals also that TFORM needs about 20\%
overhead for the Multiple Zeta Program. This is more than for programs
like Mincer.  This may be due to the use of brackets from the master
expression which may involve copious use of locks. This is still not
completely clear though.  The result is that for 8 workers the pseudo
speedup (total CPU time divided by realtime) is 7.63 while the real
speedup (compared to the FORM run) is 6.22. Of course, this is still
very good. The maximum improvement we obtained was 7.45 for a run with
17 workers.

\section{Recent development}
\begin{figure}[ht]
\begin{center}
\includegraphics[width=.75\linewidth]{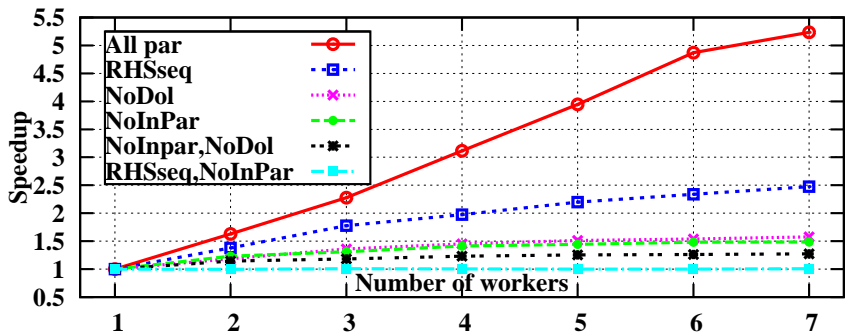}
\caption{\label{MZVtry} Results of the MZV program runs with various features
  switched off/on. The runs were for weight 20, up to depth 8.}
\end{center}
\end{figure}

Over the past years parallel FORM versions have picked up a number of
new features:
\begin{itemize}
\item {\bf Dollar variables}.
By default, both ParFORM and TFORM switch into the sequential mode
for each module which gives dollar variables a value during execution. 
But there are common cases 
when some dollar variables obtained from each term in each chunk
can be processed in parallel in order to get a minimum value, a
maximum, or a sum of results. Also, sometimes at the end of the processing
of a term the value of the dollar variable is not important at all. Hence
new module options have been implemented to help FORM to process these
variables in parallel: \verb|minimum|, \verb|maximum|, \verb|sum| and
\verb|local|.
\item {\bf Right-hand side expressions (RHS)}.
This is not a problem for TFORM since all threads work with the same
file system while it is a big problem for ParFORM since the expression
may be situated in a scratch file but different nodes may have
independent scratch file systems. For a long time ParFORM forced
evaluation of modules with RHS expressions in sequential mode. Now
ParFORM is able to perform RHS expressions in a real parallel mode.
\item {\bf InParallel statement}.
A new statement was inplemented, {\tt inparallel;}.
This statement allows the execution of complete expressions
in a single worker simultaneously. This is really useful when there
are  many short expressions, sometimes it gives a significant increase
in efficiency.
\end{itemize}
In Fig. \ref{MZVtry} we summarize the speedup curves for the TFORM
running the MZV program on 8 CPU cores computer when various features are switched off/on.
The legend is the following:
\begin{itemize} 
\item {\bf All par} -- all above mentioned features are implemented;
\item {\bf RHSseq} -- modules with RHS expressions are
  forced into the sequential mode;
\item {\bf NoDol} -- modules with dollar variables are
  forced into the sequential mode;
\item {\bf NoInar} -- no InParallel statements;
\item {\bf NoInPar,NoDol} -- modules with dollar variables
  are forced into the sequential mode, no InParallel statements;
\item {\bf RHSseq,NoInPar} modules with RHS expressions are
  forced into the sequential mode, no InParallel statements.
\end{itemize}
 As we can see, all these new features are really important.
\vspace{0.2 cm}

If FORM programs have to run for a long time the reliability of the
hardware or of the software infrastructure
becomes a critical issue. Program termination due to unforeseen
failures may waste days or weeks of invested execution time. The
checkpoint mechanism was introduced to protect long running FORM
programs as good as possible from such accidental interruptions. With
activated checkpoints FORM will save its internal state and data from
time to time on the hard disk. This data then allows a recovery from a
crash. The parallel FORM versions support this mechanism as well.

\begin{figure}[ht]
\begin{center}
\includegraphics[width=\linewidth]{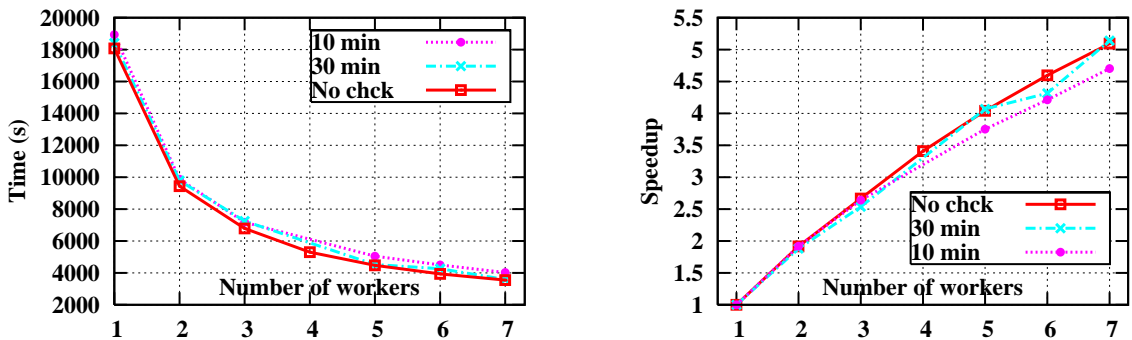}
\caption{\label{chckScal} Absolute time and speedup curves for the test program BAICER 
without checkpoint mechanism
  (``NoChck''), checkpoints every 30 minutes (``30 min'') and every 10
minutes(``10 min'').}
\end{center}
\end{figure}

By default, data are saved at the end of each module. Usually this is
too expensive. Optionally, the data may be saved only after some time
interval. The scalability for ParFORM running BAICER N=16 for
different intervals between checkpoints is depicted in Fig.~\ref{chckScal}.
As one can see, even very frequent checkpoints do not affect
performance much.

{\em Acknowledgments.}
This work was supported in part by DFG through SBF/TR 9 and by the FOM 
foundation.

\end{document}